\documentclass[secnumarabic,amssymb, nobibnotes, aps, prd]{revtex4}
\usepackage{amsmath} \usepackage{amssymb}
\usepackage{graphicx} %use graph format
\usepackage{epstopdf}
\usepackage{xcolor}
\usepackage{amsmath}
\usepackage{amsmath,amssymb,amsthm,amsfonts,mathrsfs,bm,verbatim}
\usepackage{graphicx,subfigure}
\usepackage{indentfirst}
\usepackage[backref]{hyperref}

\newcommand{\bea}{\begin{eqnarray}}
\newcommand{\eea}{\end{eqnarray}}

\begin{document}

\title{Quantum transport theory with vector interaction}
\author{Peiwei Yu}
\affiliation{Guangdong Provincial Key Laboratory of Nuclear Science, Institute of Quantum Matter, 
South China Normal University, Guangzhou 510006, China.} 
\affiliation{Guangdong-Hong Kong Joint Laboratory of Quantum Matter, South China Normal University, Guangzhou 510006, China.}

\author{Xingyu Guo}
\email{guoxy@m.scnu.edu.cn}
\affiliation{Guangdong Provincial Key Laboratory of Nuclear Science, Institute of Quantum Matter, 
South China Normal University, Guangzhou 510006, China.} 
\affiliation{Guangdong-Hong Kong Joint Laboratory of Quantum Matter, South China Normal University, Guangzhou 510006, China.}

\begin{abstract}
We derive the relativistic quantum kinetic equation for massless fermions with vector and axial vector interaction using the Wigner function formalism. The vector and axial vector currents are self-consistently treated with corresponding constraint equations. The kinetic equations are derived and the condition for equilibrium is discussed up to the first order of $\hbar$. In addition to the vorticity and shear contributions, the divergence of the local particle number density is also found to contribute to the kinetic equations.
\end{abstract}

\maketitle

\section{Introduction}
Quantum chromodynamics(QCD) predicts the existence of a new state of matter, the quark-gluon plasma(QGP). It is widely accepted that QGP is created in relativistic heavy-ion collisions create \cite{STAR:2021beb,STAR:2019fge,CMS:2018zza}. In the QGP, the chiral symmetry is restored and light quarks can be considered massless fermions. On the other hand, there are local chiral imbalances due to triangle anomaly. The interaction between the chiral anomaly and the magnetic field or vorticity gives rise to novel transport phenomena such as the chiral magnetic effect(CME)\cite{Kharzeev:2007jp} and the chiral vortical effect(CVE)\cite{Son:2009tf}. The vorticity will also result in the polarization of particles which is confirmed by the polarization of the $\Lambda$ hyperion\cite{STAR:2023nqc,Gou:2023dkp,Sarkar:2022uyt}. In such a system, the chirality and spin degree of freedom are important. Also, the study of local $\Lambda$ polarization shows that the non-equilibrium effect should be considered\cite{Gou:2023dkp,Sarkar:2022uyt}. The chiral kinetic theory is among the effective theoretical tools to study such topics. Many results were obtained, such as the anomalous chiral transport equation in heavy-ion collisions\cite{Son:2012zy,Gao:2012ix,Chen:2012ca,Huang:2018wdl,Wang:2020dws}, the quantum kinetic theory for massive fermions under external fields\cite{Chen:2013iga,Gao:2019znl,Weickgenannt:2019dks,Hattori:2019ahi} and the non-relativistic kinetic of spin-polarized system\cite{Zamanian:2009jzf}.

Previous works show that vorticity and shear play important roles in the system\cite{Lin:2022tma,Liu:2022zxd,Buzzegoli:2022fxu,Becattini:2021suc}. They can be introduced as $\hbar$ order corrections to the free-streaming part of the kinetic equations. On the other hand, when fermion-fermion interaction is taken into consideration, effective vector and axial vector interaction terms appear in the collision part. It is also necessary to check the self-consistency of the vorticity and shear, as they come from the velocities of the fermion field itself. In the past, many works focused on the Nambu–Jona-Lasinio (NJL) model, such as the spin polarization for massive fermion\cite{Wang:2021owk} and the collision term\cite{Wang:2020pej}. But to better study the spin-related phenomenon, it might be helpful to directly consider an effective vector and axial vector interaction. Therefore, in this work, we study the relativistic quantum kinetic equation for massless fermions with effective vector axial interaction. This Lagrangian goes back to Fermi's theory of weak interaction\cite{PhysRev.109.193}. In QCD, the vector interaction can be used to explain the one-gluon exchange\cite{Fukushima:2010zza} and is used as an effective theory for 4-fermion interaction with axial interaction\cite{Diakonov:2011fs,Andrianov:2002iw}. Furthermore, the study of vector interaction and axial interaction profit for the derivation of anomalous hydrodynamics\cite{Satow:2014lia,Gorbar:2017toh,Buzzegoli:2017cqy,Buzzegoli:2018wpy}, magnetohydrodynamics\cite{Lin:2021sjw} and spin hydrodynamics\cite{Florkowski:2018fap,Bhadury:2020puc,Peng:2021ago}. The mean-field approximation is used for simplicity, and the vector and axial vector currents are self-consistently constrained by constraint equations. The semi-classical expansion is taken up to the $\hbar$ order.In the zero-order. We get the on-shell conditions, the kinetic equations, and the equilibrium conditions for left- and right-handed components.

The paper is organized as follows: In Sect.2, we derive the equation of motion for the Wigner function without the collision term. In Sect.3, the $\hbar$ expansion is used to simplify the equations. Finally, we make a brief discussion and conclusion in Sect,4.

\section{EQUATION OF MOTION FOR THE WIGNER FUNCTION}
We start from a Lagrangian density with effective vector and axial vector interaction terms\cite{PhysRev.109.193}
\begin{align}
\mathcal{L}=  i\gamma ^{\mu }\overline{\psi }\partial_{\mu } \psi +G\left [ \left ( \overline{\psi }\gamma ^{\mu }\psi \right )^{2}+\left ( \overline{\psi }\gamma _{5}\gamma ^{\mu }\psi \right )^{2} \right ]. \label{eq1}
\end{align}
With the mean-field approximation $(\bar{\psi}\gamma^\mu\psi)^2=\langle\bar{\psi}\gamma^\mu\psi\rangle \bar{\psi}\gamma^\mu\psi$ and $(\bar{\psi}i\gamma_5\gamma^\mu\psi)^2=\langle\bar{\psi}\gamma^\mu\gamma_5\psi\rangle \bar{\psi}\gamma_5\gamma^\mu\psi$, the Lagrangian density can be reduced as
\begin{align}
\mathcal{L}=\overline{\psi }\left ( i\gamma ^{\mu }\partial_{\mu }+\gamma ^{\mu }J_{V_{\mu }}+\gamma _{5}\gamma ^{\mu }J_{A_{\mu }} \right )\psi, \label{eq2}
\end{align}
where $J_{V}^{\mu }=G\left<\overline{\psi }\gamma ^{\mu }\psi \right>$ and $J_{A}^{\mu }=G\left<\overline{\psi }\gamma _{5}\gamma ^{\mu }\psi \right>$. Here $J_V$ and $J_A$ are not to be considered as condensates, but as the vector and axial vector currents of the system that we are interested in. For example, in heavy-ion collisions, the fireball is rapidly expanding, resulting in a radial divergence and current. What is more, the large angular momentum in non-central collisions means there is a modification to the current that has non-zero derivatives. The chiral anomaly, along with possible CME and CVE effects will also give an axial vector current.  As the fireball is a thermal system, these expectation values can be treated as ensemble averages. For simplicity, the coupling constant is absorbed into the currents.

From the mean-field effective Lagrangian density, one can obtain the Dirac equation:
\begin{align}
\overline{\psi }\left ( i\gamma ^{\mu }\overleftarrow{\partial_{\mu } }-\gamma ^{\mu }J_{V_{\mu }}-\gamma _{5}\gamma ^{\mu }J_{A_{\mu }} \right )=0,  \nonumber\\
\left ( i\gamma ^{\mu }\overrightarrow{\partial_{\mu }}+\gamma ^{\mu }J_{V_{\mu }}+\gamma _{5}\gamma ^{\mu }J_{A_{\mu }} \right )\psi =0. \label{eq3}
\end{align}
The Wigner function for fermions is
\begin{align}
W\left ( x,p \right )=\int d^{4}y e^{ipy}\left<\psi \left ( x+\frac{y}{2} \right )\overline{\psi }\left ( x-\frac{y}{2} \right ) \right>,  \label{eq4}
\end{align}
where we did not consider the effect of gauge fields. Combining the above Dirac equations and the Wigner function, one can get the equation of motion for the Wigner function\cite{Huang:2018wdl, Lin:2022tma}
\begin{align}
\left ( \gamma ^{\mu }K_{\mu }-\gamma _{5}\gamma ^{\mu }K_{5\mu } \right )W =0, \label{eq5}
\end{align}
where
\begin{align}
K_{\mu} =&\Pi_{\mu } +iD_{\mu }, \nonumber \\
K_{5\mu} =&\Pi_{5\mu } +iD_{5\mu }, \nonumber \\
\Pi_{\mu } =&p_{\mu }+cos\left ( \frac{\hbar }{2}\nabla  \right )J_{V_{\mu}}, \nonumber \\
D_{\mu } =&\frac{ \hbar}{2}\partial _{\mu }-sin\left ( \frac{\hbar }{2}\nabla  \right )J_{V_{\mu}}, \nonumber \\
\Pi_{5\mu } =&-cos\left ( \frac{\hbar }{2}\nabla  \right )J_{A_{\mu}}, \nonumber\\
D_{5\mu} =&sin\left ( \frac{\hbar }{2}\nabla  \right )J_{A_{\mu}}, \nonumber \\
\nabla =&\partial_{x} \cdot \partial _{p}. \nonumber
\end{align}
The operator $\bigtriangledown$ is $\partial_{x} \cdot \partial _{p}$, in which the spatial derivative $\partial_{x}$ only acts on $J_{A_{\mu}}$ or $J_{V_{\mu}}$, but not on the Wigner function $W$.

The Wigner function is a $4 \times 4$ matrix and satisfies the relationship $\gamma_{0}W^{+}\gamma_{0}=W$. It can be decomposed in terms of 16 independent generators of Clifford algebra\cite{Huang:2018wdl}. So the equation of motion can be decomposed to 10 equations\cite{Guo:2017dzf}
\begin{align}
\Pi_{\mu}V^{\mu}+\Pi_{5\mu}A^{\mu}=0, \nonumber\\
D_{\mu}A^{\mu}+D_{5\mu}V^{\mu}=0, \nonumber\\
\Pi_{\mu}F+D^{\nu}S_{\mu\nu}-D_{5\mu}P+\frac{1}{2}\varepsilon _{\mu \nu \sigma \rho }\Pi^{\nu}_{5}S^{\sigma \rho}=0, \nonumber\\
D_{\mu}P-\frac{1}{2}\varepsilon _{\mu \nu \sigma \rho }\Pi^{\nu}S^{\sigma \rho}-\Pi_{5\mu}F-D^{\nu}_{5}S_{\mu\nu}=0,\nonumber\\
D_{\mu}V_{\nu}-D_{\nu}V_{\mu}-\varepsilon _{\mu \nu \sigma \rho }\Pi^{\sigma}A^{\rho}-\varepsilon _{\mu \nu \sigma \rho }\Pi^{\sigma}_{5}V^{\rho}+D_{5\mu}A_{\nu}-D_{5\nu}A_{\mu}=0, \nonumber\\
D_{\mu}V^{\mu}+D_{5\mu}A^{\mu}=0, \nonumber\\
\Pi_{\mu}A^{\mu}+\Pi_{5\mu}V^{\mu}=0, \nonumber\\
D_{\mu}F-\Pi^{\nu}S_{\mu\nu}+\Pi_{5\mu}P+\frac{1}{2}\varepsilon _{\mu \nu \sigma \rho }D^{\nu}_{5}S^{\sigma\rho}=0, \nonumber\\
\Pi_{\mu}P+\frac{1}{2}\varepsilon _{\mu \nu \sigma \rho }D^{\nu}S^{\sigma\rho}+D_{5\mu}F-\Pi^{\nu}_{5}S_{\mu\nu}=0, \nonumber\\
\Pi_{\mu}V_{\nu}-\Pi_{\nu}V_{\mu}+\varepsilon _{\mu \nu \sigma \rho }D^{\sigma}A^{\rho}+\varepsilon _{\mu \nu \sigma \rho }D^{\sigma}_{5}V^{\rho}+\Pi_{5\mu}A_{\nu}-\Pi_{5\nu}A_{\mu}=0.  \label{eq6}
\end{align}
Where $V_\mu$, $A_\mu$, $F$, $P$, $S_{\mu\nu}$ are the vector, axial vector, scalar, pseudoscalar, and tensor components respectively. $\epsilon_{\mu\nu\sigma\rho}$ is the Levi-Civita symbol.

The roles of the currents look similar to gauge fields. But they are different. In some studies\cite{Huang:2018wdl,Guo:2020zpa}, the gauge fields are taken as classical and their strength is decided beforehand. In other ones\cite{Lin:2022tma}\cite{Lin:2021mvw}, they are treated as dynamical degrees of freedom, with their own transport equations. In both cases, the gauge fields are not determined by the fermion field. On the contrary, the currents $J_V$ and $J_A$ are by definition part of the fermion field, and they are connected to the vector and axial-vector components.
\begin{eqnarray}
J_V^\mu(x) &=G\int d^4 p V^\mu(x,p), \label{eq7}\\
J_A^\mu(x)&=G\int d^4 p A^\mu(x,p).  \label{eq8}
\end{eqnarray}

Similar to the usual constraint equations that determine the value of sigma and pion condensates\cite{Wang:2021owk, Huang:2020wrr,Wang:2017vtn}, these equations can be used to determine the possible value of particle number density and axial imbalance, which we will discuss later.

It is simpler to consider the chiral components
\begin{align}
V_{\chi \mu}=V_{\mu}+\chi A_{\mu}, \label{eq9}
\end{align}
where $\chi=\pm1$ corresponds to right-handed and left-handed components. The equation of motion for $V_{\chi\mu}$ are
\begin{align}
\left( \Pi _{\mu }+\chi \Pi _{5\mu } \right) V_{\chi }^{\mu }=0, \nonumber\\
\left( D_{\mu} + \chi D_{5\mu} \right) V_{\chi}^{\mu}=0, \nonumber\\
\left( \Pi_{\mu} +\chi \Pi_{5\mu} \right) V_{\chi\nu} - \left( \Pi_{\nu}+\chi \Pi_{5\nu} \right) V_{\chi\mu}+\chi \varepsilon_{\mu \nu \sigma \rho } \left ( D^{\sigma }+\chi D_{5}^{\sigma } \right )V_{\chi }^{\rho }=0. \label{eq10}
\end{align}
One can see that the chiral components are decoupled from each other as well as the other components.

\section{Transport equation}
The equations are still difficult to solve, but the operators and functions can be expanded by $\hbar$ and solved order by order \cite{Guo:2020zpa}.

\subsection{The zeroth order}
To the zeroth order,  eq.(\ref{eq6}) can be written as
\begin{align}
\left ( p_{\mu}+J^{\left ( 0 \right )}_{V_{\mu}} \right )F^{\left ( 0 \right )}-\frac{1}{2}\varepsilon _{\mu \nu \sigma \rho }J^{\left ( 0 \right )\nu}_{A}S^{\left ( 0 \right )\sigma\rho}=0 \nonumber\\
-\frac{1}{2}\varepsilon _{\mu \nu \sigma \rho }\left ( p^{\nu}+J_{V}^{\left ( 0 \right )\nu} \right )S^{\left ( 0 \right )\sigma\rho}+J^{\left ( 0 \right )}_{A_{\mu}}F^{\left ( 0 \right )}=0 \nonumber\\
-\left ( p^{\nu}+J_{V}^{\left ( 0 \right )\nu} \right )S^{\left ( 0 \right )}_{\mu\nu}-J^{\left ( 0 \right )}_{A_{\mu}}P^{\left ( 0 \right )}=0 \nonumber\\
\left ( p_{\mu}+J^{\left ( 0 \right )}_{V_{\mu}} \right )P^{\left ( 0 \right )}+J^{\left ( 0 \right )\nu}_{A}S^{\left ( 0 \right )}_{\mu\nu}=0 \label{eq11}
\end{align}
Form eq.(\ref{eq11}), we can get the on-shell conditions
\begin{align}
\left(\left( p+J_{V}^{\left ( 0 \right )} \right)^{2}+J^{\left ( 0 \right )2}_{A}\right)F^{\left(0\right)}=0, \label{eq12}\\
\left(\left( p+J_{V}^{\left ( 0 \right )} \right)^{2}+J^{\left ( 0 \right )2}_{A}\right)P^{\left(0\right)}=0, \label{eq13}\\
\left(\left( p+J_{V}^{\left ( 0 \right )} \right)^{2}+J^{\left ( 0 \right )2}_{A}\right)S^{\left( 0 \right) \mu\nu}=0, \label{eq14}\\
\left ( p+J_{V}^{\left ( 0 \right )}+\chi J_{A}^{\left ( 0 \right )} \right )^{2}V^{\left ( 0 \right )}_{\chi\nu}=0 . \label{eq15}
\end{align}
Form eq.(\ref{eq15}), we find that even at the 0th order, the on-shell conditions are modified, and they are different for left-hand, right-hand, and other components. This can be viewed as a shift of masses and/or as the modification of mechanical momenta.
The tensor components can be expressed by the scalar and pseudoscalar components
\begin{align}
&S^{\left( 0 \right) \sigma\rho}=\frac{1}{J_{A}^{\left ( 0 \right )2}}\left( \varepsilon ^{\mu \nu \sigma \rho }\left( p_{\mu}+J_{V_{\mu}}^{\left ( 0 \right )} \right)J_{A_{\nu}}^{\left ( 0 \right )}F^{\left(0\right)}-\left( p^{\rho}+J_{V}^{\left ( 0 \right )\rho}\right)J_{A}^{\left ( 0 \right )\sigma}P^{\left(0\right)}+\left( p^{\sigma}+J_{V}^{\left ( 0 \right )\sigma}\right)J_{A}^{\left ( 0 \right )\rho}P^{\left(0\right)} \right) \label{eq16},
\end{align}
So among them, the degree of freedom is 2.

We will then focus on the chiral components, the kinetic equation for chiral components are
\begin{align}
\widetilde{p}_{\chi \mu }V^{\left ( 0 \right )\mu}_{\chi}=0, \nonumber\\
\widetilde{p}_{\chi \mu }V^{\left ( 0 \right )}_{\chi\nu}-\widetilde{p}_{\chi \nu }V^{\left ( 0 \right )}_{\chi\mu}=0, \nonumber\\
\nabla_{\chi \mu } V^{\left ( 0 \right )\mu}_{\chi}=0. \label{eq17}
\end{align}
where $\widetilde{p}_{\chi} = p+J_\chi$, $J_\chi = J_V+\chi J_A$, $\nabla_\chi\mu=\partial_{\beta}J^{\left(0\right)}_{\chi\mu}\partial^{\beta}_{p}$
and the constraint equation for the chiral currents is
\begin{align}
J_{\chi \mu }&=G\int d^{4}p V_{\chi \mu } \label{eq18}.
\end{align}

From these equations, the chiral components  can be expressed as $V^{\left ( 0 \right )}_{\chi\mu}=\widetilde{p}_{\chi \mu }f_{\chi }^{\left ( 0 \right )}\delta \left ( \widetilde{p}_{\chi }^{2} \right )$, Putting this into the transport equation we get
\begin{align}
\widetilde{p}_{\chi }^{\mu }\left ( \nabla_{\chi \mu }f_{\chi }^{\left ( 0 \right )}  \right )\delta \left ( \widetilde{p}_{\chi }^{2} \right )=0. \label{eq19}
\end{align}

This is very similar to the free-streaming transport equation. The only difference is that the momentum is shifted.

We assume the equilibrium distribution as
\begin{align}
f^{\left ( 0 \right )}_{\chi~eq}=\frac{1}{e^{\frac{|p \cdot u_\chi|}{T}}+1}, \label{eq20}
\end{align}
where $T$ is temperature and $u_{\chi\mu}=\frac{J^{(0)}_{\chi\mu}}{G n_\chi}$ is the fluid velocity, $n_\chi = \frac{1}{G}\sqrt{J^{(0)\mu}_\chi J^{(0)}_{\chi\mu}}\mathrm{sgn}(J^{(0)0}_\chi)$ is the particle number density. Putting eq.(\ref{eq20}) into eq.(\ref{eq19}), we can get the equilibrium condition at zeroth order
\begin{align}
(p^{\mu }p^{\nu }+J_{\chi }^{( 0)\mu }J_{\chi }^{(0)\nu})\left[ \partial_{\mu}(\frac{u_{\chi\nu}}{T})+\partial_{\nu}(\frac{u_{\chi\mu}}{T})\right]+(2J^{(0)}_\chi-p)\cdot\partial(\frac{n_\chi}{T})=0. \label{eq21}
\end{align}

We can see that in general, the thermal shear tensor $\partial_\mu(\frac{u_\nu}{T})+\partial_\nu(\frac{u_\mu}{T})$ is not required to be 0. Only when the particle number density is a constant do we get the usual killing condition.

Finally, we consider the constraint equation. Eq.(\ref{eq17}) and eq.(\ref{eq18}) give
\begin{align}
J_{\chi\mu}^{\left(0\right)}=G \int d^{4}p \widetilde{p}_{\chi\mu}f_{\chi}^{\left(0\right)} \delta \left(\widetilde{p}^{2}_{\chi}\right). \label{eq22}
\end{align}

\begin{figure}[!hbt]
\begin{center}
\includegraphics[width=0.6\textwidth]{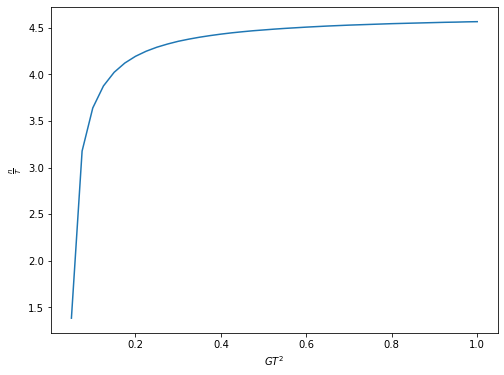}
\end{center}
\label{fig:n-T}
\caption{The $|\frac{n}{T}|$ as a function of $GT^2$}
\end{figure}

This gives constrain on the current. Especially, if we consider the equilibrium distribution, we have

\begin{align}
\frac{n}{T}=\frac{2}{3}\pi^{3}\frac{n}{T}GT^{2}+\left[4 \ln2\cdot \pi \frac{n^2}{T^2}GT^2+6\pi \zeta\left ( 3 \right )GT^{2}+4 \pi GT^{2} Li_{3}\left ( -e^{\frac{n}{T}} \right )+4 \pi GT^{2}Li_{3}\left ( -e^{-\frac{n}{T}} \right )\right]\mathrm{sgn}(n), \label{eq23}
\end{align}

where $Li_{s}\left(x\right)$ is the polylogarithm function\cite{Gorbar:2017awz} and $\zeta$ is the Riemann zeta function. The chiral index is omitted here because the equation does not depend on chirality, even though the chiral currents can be different. This is because the interaction is symmetric for left- and right-handed components. The quantity $\frac{n}{T}$ can be viewed as a dimensionless parameter determined by $GT^2$. Performing Taylor expansion, we get
\begin{align}
\frac{n}{T}\approx GT^2(\frac{2}{3}\pi^3\frac{n}{T}-\frac{\pi}{12}\frac{n^4}{T^4}\mathrm{sgn}(n)). \label{eq24}
\end{align}
Therefore when $GT^2<\frac{3}{2\pi^3}\approx0.05$, the constraint equation has only trivial solution $n=0$. When $GT^2\ge0.05$, there are two additional opposite non-zero solutions whose absolute value is shown in Fig.\ref{fig:n-T}

This indicates that without chemical potential, the vector interaction can result in a non-zero net local chiral particle number density at finite temperatures, with equal possibilities of being positive and negative. Therefore in a statistical many-body system with zero chemical potential, these non-zero solutions will serve as local fluctuations, and the expectation value of particle density remains zero.

\subsection{The $\hbar$ order}
From the $\hbar$ order of the kinetic equation for chiral components, we can get the solution of $V^{(1)}_{\chi\mu}$
\begin{align}
V^{(1)}_{\chi \mu }&=\widetilde{p}_{\chi \mu }f^{(1)}_{\chi }\delta \left ( \widetilde{p}^{2}_{\chi } \right )-\frac{\chi  }{2\left ( \widetilde{p}_{\chi }\cdot u' \right )}\varepsilon _{\mu \nu \alpha \sigma }u^{'\nu }\widetilde{p}_{\chi }^{\alpha }\left ( \nabla ^{\sigma }_{\chi }f_{\chi }^{\left ( 0 \right )} \right )\delta \left ( \widetilde{p}_{\chi }^{2} \right ) \nonumber\\
&+\frac{\chi  }{\widetilde{p}^{2}_{\chi }}\varepsilon _{\mu \nu \sigma \rho }\widetilde{p}_{\chi }^{\nu }\Omega_{\chi } ^{\sigma \rho }f^{\left ( 0 \right )}_{\chi }\delta \left ( \widetilde{p}_{\chi }^{2} \right )-\frac{2\widetilde{p}_{\chi \mu }\widetilde{p}_{\chi \nu }J_{\chi  }^{\left ( 1 \right )\nu }}{\widetilde{p}_{\chi }^{2}}f_{\chi }^{\left ( 0 \right )}\delta \left ( \widetilde{p}^{2}_{\chi } \right )+ J_{\chi \mu }^{\left ( 1 \right )}f_{\chi }^{\left ( 0 \right )}\delta \left ( \widetilde{p}_{\chi }^{2} \right ), \label{eq25}
\end{align}
where $\bigtriangledown_{\chi \mu }=\partial _{\mu }-\bigtriangledown J_{\chi\mu }^{\left ( 0 \right )}$, $\Omega_{\chi \mu \nu } =\frac{1}{2}\left ( \partial _{\mu }J_{\chi \nu }^{\left ( 0 \right )} -\partial _{\nu }J_{\chi \mu }^{\left ( 0 \right )} \right )=\frac{n_\chi}{2}(\partial_\mu u_{\chi\nu}-\partial_\nu u_{\chi\mu})+\frac{1}{2}(u_{\chi\nu}\partial_\mu-u_{\chi\mu}\partial_\nu)n_\chi$ and $u'$ is an additional time-like vector. One should notice that here the kinetic vorticity instead of the thermal vorticity tensor appears because we have not considered the equilibrium condition. The detailed calculation is in Appendix B. 

In the following derivation, we will choose $u'$ to be $J^{\left(0\right)}_{\chi\mu}$.
According to eq.(\ref{eq25}), we can get the transport equation for $f_{\chi}=f_{\chi}^{(0)}+\hbar f_{\chi}^{(1)}$
\begin{align}
&\delta\left ( \widetilde{p}_{\chi }^{2}+\frac{2\chi \hbar}{\widetilde{p}_{\chi }\cdot J_{\chi}^{\left ( 0 \right )}}\widetilde{p}_{\chi }^{\alpha }\widetilde{\Omega}_{\chi \alpha \nu }J_{\chi}^{\left ( 0 \right )\nu }+2\hbar\widetilde{p}_{\chi \nu }J_{\chi }^{\left ( 1 \right )\nu } \right ) \{ \widetilde{p}_{\chi }\cdot \nabla _{\chi } \nonumber\\
&+\frac{\chi \hbar}{2\left ( \widetilde{p}_{\chi }\cdot J_{\chi}^{\left ( 0 \right )} \right )^{2}}\left [ \left ( \partial ^{\mu }J_{\chi}^{\left ( 0 \right )\beta }\right)\widetilde{p}_{\chi \beta }-\Omega_{\chi }^{\beta \mu }J_{\chi \beta}^{\left ( 0 \right )}  \right ]\varepsilon _{\mu \nu \alpha \sigma }J_{\chi}^{\left ( 0 \right )\nu }\widetilde{p}_{\chi }^{\alpha } \nabla _{\chi }^{\sigma }\nonumber\\
&-\frac{\chi \hbar}{\widetilde{p}_{\chi }\cdot J_{\chi}^{\left ( 0 \right )}}\widetilde{\Omega }_{\chi \alpha \sigma  } \widetilde{p}_{\chi }^{\alpha }\nabla _{\chi }^{\sigma }+\frac{\chi \hbar}{2\widetilde{p}_{\chi }\cdot J_{\chi}^{\left ( 0 \right )}}J_{\chi\nu}^{\left ( 0 \right )}\widetilde{p}_{\chi \alpha}\left ( \partial ^{\beta }\widetilde{\Omega}_{\chi }^{\nu \alpha } \right )\partial _{\beta }^{p}+\hbar J_{\chi\mu}^{\left ( 1 \right )}\bigtriangledown _{\chi }^{\mu }\nonumber\\
&-\frac{\hbar}{\widetilde{p}_{\chi }^{2}}\widetilde{p}_{\chi \delta }\Omega_{\chi }^{\mu \delta }J_{\chi \mu }^{\left ( 1 \right )} -\hbar\partial^{\beta } J_{\chi }^{\left ( 1 \right )\mu }\widetilde{p}_{\chi \mu }\partial ^{p}_{\beta } \}f_{\chi}=0. \label{eq26}
\end{align}
Where $\widetilde\Omega_{\chi\mu\nu} = \frac{1}{2}\varepsilon_{\mu\nu\sigma\rho}\Omega^{\sigma\rho}_\chi$. The derivation of this equation will be shown in Appendix C\cite{Guo:2020zpa}. From eq.(\ref{eq26}), the mass shell is shifted again at the $\hbar$ order by the momentum translation and vorticity.

Then we consider a special condition that is similar to the situation in heavy-ion collisions. In the fireball, chiral currents only come from fluctuations, so we can safely assume $J_{A\mu}^{(0)}$ is zero. In that case, up to the 0th order, $J_+^{(0)} = J_-^{(0)}=J$ and $f_-^{(0)}=f_+^{(0)}=f^{(0)}$, with the same mass shell. For simplicity, we will also assume $J^{(1)}_V$ to be zero. So $V_{\chi\mu}$  become
\begin{align}
V_{\chi \mu }&=\widetilde{p}_{\mu }f_{\chi}\delta \left ( \widetilde{p}^{2} \right )-\frac{\chi \hbar }{2\widetilde{p}\cdot u}\varepsilon _{\mu \nu \alpha \sigma }u^{\nu}\widetilde{p}^{\alpha }\left ( \nabla^{\sigma }f^{\left ( 0 \right )} \right )\delta \left ( \widetilde{p}^{2} \right )+\frac{\chi \hbar }{\widetilde{p}^{2}}\varepsilon _{\mu \nu \sigma \rho }\widetilde{p}^{\nu }\Omega ^{\sigma \rho }f^{\left ( 0 \right )}\delta \left ( \widetilde{p}^{2} \right ) \label{26}\\
V_{\mu}&=\frac{1}{2} \left( V_{+}+V_{-} \right)=\widetilde{p}_{\mu }f^{(0)}\delta \left ( \widetilde{p}^{2} \right )\label{27}\\
A_{\mu}&=\frac{1}{2} \left(V_{+}-V_{-}\right)=\frac{1}{2}\widetilde{p}_{\mu }\left ( f_{+}^{(1)}-f_{-}^{(1)} \right )\delta \left ( \widetilde{p}^{2} \right )-\frac{\hbar }{2\widetilde{p}\cdot u}\varepsilon _{\mu \nu \alpha \sigma }u^{\nu}\widetilde{p}^{\alpha }\left ( \nabla^{\sigma }f^{\left ( 0 \right )} \right )\delta \left ( \widetilde{p}^{2} \right )+\frac{\hbar }{\widetilde{p}^{2}}\varepsilon _{\mu \nu \sigma \rho }\widetilde{p}^{\nu }\Omega ^{\sigma \rho }f^{\left ( 0 \right )}\delta \left ( \widetilde{p}^{2} \right )  \label{28}
\end{align}
For all components that become independent of chirality, the chiral indices $\chi$ are dropped. Eq.(\ref{28}) indicates that spin polarization can be induced by vector and axial vector interaction.  In addition to the usual chiral vortical effect, there is a contribution from the divergence of the number density. To show this, we set the vorticity tensor to zero, then
\begin{align}
A_{\mu}&=\frac{1}{2}\widetilde{p}_{\mu }\left ( f_{+}^{(1)}-f_{-}^{(1)} \right )\delta \left ( \widetilde{p}^{2} \right )-\frac{\hbar }{2\widetilde{p}\cdot u}\varepsilon _{\mu \nu \alpha \sigma }u^{\nu}\widetilde{p}^{\alpha }\left ( \nabla^{\sigma }f^{\left ( 0 \right )} \right )\delta \left ( \widetilde{p}^{2} \right )+\frac{\hbar }{\widetilde{p}^{2}}\varepsilon _{\mu \nu \sigma \rho }\widetilde{p}^{\nu }u^{\rho}\left ( \partial ^{\sigma }n \right )f^{\left ( 0 \right )}\delta \left ( \widetilde{p}^{2} \right )  \label{29} 
\end{align}
As mentioned above, in the fireball, the gradient of the number density is in the radial direction, so we will have an axial vector current in the angular direction from the last term. This indicates that there is a local spin distribution, which is qualitatively consistent with experimental findings\cite{Wu:2022mkr,Sarkar:2022uyt,Niida:2018hfw}. And combining this with eq.\ref{eq21} one can see that the polarization is related to the shear tensor.

There are also contributions to the transport equations
\begin{align}
\nabla ^{\mu }V_{\mu}&=\widetilde{p}_{\mu } \nabla^{\mu } f^{(0)}\delta \left ( \widetilde{p}^{2} \right )=0\label{31}\\
\nabla ^{\mu} A_{\mu}&=\frac{1}{2}\widetilde{p}_{\mu }\left ( \nabla^{\mu } f_{+}^{(1)}-\nabla^{\mu } f_{-}^{(1)}\right )\delta \left ( \widetilde{p}^{2} \right )-\frac{\hbar}{\widetilde{p}\cdot J_{V}^{\left ( 0 \right )}}\varepsilon _{\mu \nu \alpha \rho  }u^{\rho  }\left ( \partial^{\mu } n  \right )J_{V}^{\left ( 0 \right )\nu   }\widetilde{p}^{\alpha }\widetilde{p}_{\sigma }\left ( \nabla^{\sigma }f^{\left ( 0 \right )} \right )\delta '\left ( \widetilde{p}^{2} \right ) \nonumber\\
&+\frac{\hbar}{2\left ( \widetilde{p}\cdot J_{V}^{\left ( 0 \right )} \right )^{2}}\left [\left ( \partial ^{\mu }J_{V}^{\left ( 0 \right )\beta } \right )\widetilde{p}_{\beta  }-\frac{1}{2}J^{\left ( 0 \right )}_{V_{\beta  } }\left ( u^{\mu }\partial ^{\beta  }-u^{\beta }\partial ^{\mu } \right )n \right ]\varepsilon _{\mu \nu \alpha  \sigma  }J_{V}^{\left ( 0 \right )\nu }\widetilde{p}^{\alpha }\left ( \nabla^{\sigma }f^{\left ( 0 \right )} \right )\delta \left ( \widetilde{p}^{2} \right ) \nonumber\\
&+\frac{\hbar}{2\widetilde{p}\cdot J_{V}^{\left ( 0 \right )} }\varepsilon _{\mu \nu \alpha \sigma } \partial ^{\beta }\left [ u^{\nu }\left ( \partial ^{\sigma  }n \right ) \right ]J_{V}^{\left ( 0 \right )\nu }\widetilde{p}^{\alpha }\left ( \partial _{\beta }^{p}f^{\left ( 0 \right )} \right )\delta \left ( \widetilde{p}^{2} \right ) \nonumber\\
&=0 \label{32}
\end{align}

There should also be a constraint equation for the 1st-order currents
\begin{align}
J^{(1)}_{\chi\mu}=\int d^4 p V^{(1)}_{\chi\mu}. \label{eq30}
\end{align}
But we do not discuss the possible form of the 1st order distribution function $f^{(1)}_\chi$ in this study, so the constraint equation is not solved.

\section{Summary}
In this work, we derive the relativistic quantum kinetic equation for massless fermions with vector and axial interaction from the Wigner function formalism and carry out the semi-classical expansion up to $\hbar$ order. Our results show that there is a non-trivial mass-shell shift and non-zero particle number density even at the 0th order of $\hbar$. We also show that there can be a non-zero shear tensor and gradient of particle number density in the equilibrium state, and they will contribute to the transport equation. There will be additional contributions to the axial current or the spin polarization. Therefore, the effective vector interaction could be useful in the study of spin polarization in heavy-ion collisions. One can further research the spin polarization which is produced by the vector interaction and axial interaction. For massless Fermions, the direction of spin is not well-defined, but we can find the direction of spin polarization by chirality. Also, as the interaction term caused the mass shell to be shifted, it is possible that now the fermions can be treated as massive, then methods such as the Matrix Valued Spin dependent Distribution can be used\cite{Hidaka:2022dmn}.

\section*{Acknowledgments}
This work is supported by Guangdong Major Project of Basic and Applied Basic Research No.2020B0301030008, Science and Technology Program of Guangzhou No. 2019050001, and NSFC grant Nos. 11905006 and 12035007.

\appendix

\section{The polylogarithm function}
In this appendix, we present some formulas used in the calculation of eq.(\ref{eq23}).
Let us introduce the following integral
\begin{align}
\int_{0}^{\infty }dp p^{k}\frac{1}{1+e^{\left ( vp\mp \mu  \right )/T}}=-\frac{T^{k+1}\Gamma \left ( k+1 \right )}{v^{k+1}}Li_{k+1} \left ( -e^{\pm \mu /T} \right ) ,   \label{a1}
\end{align}
where $\Gamma \left ( n \right )$ is the gamma function and $Li_{n+1} \left( x \right)$ is the polylogarithm function. It is easy to verify that the RHS of eq.(\ref{eq22}) is an off function of $n$, so we only consider the $n\ge 0$ situation. Assuming $n>0$, eq.(\ref{eq22}) can be written as
\begin{align}
n&=G\int d^{4}p E_{\chi } \frac{1}{e^{\frac{E_{\chi }-n}{T}}+1}\delta \left ( \widetilde{p}_{\chi } \right ) \nonumber\\
&=2 \pi G \int_{0}^{n} dp \frac{p^{2}}{e^{-\frac{p-n}{T}}+1} +2 \pi G \int_{n}^{\infty } dp \frac{p^{2}}{e^{\frac{p-n}{T}}+1} - 2 \pi G \int_{0}^{\infty } dp \frac{p^{2}}{e^{\frac{p+n}{T}}+1}, \label{a4}
\end{align}
where
\begin{align}
2 \pi G \int_{0}^{n} dp \frac{p^{2}}{e^{-\frac{p-n}{T}}+1} &=\frac{1}{3}\pi^{3}nGT^{2}+2 ln2\cdot \pi n^{2}GT+3\pi \zeta\left ( 3 \right )GT^{3}+4 \pi GT^{3}Li_{3}\left ( -e^{\frac{n}{T}} \right ), \nonumber\\
2 \pi G \int_{n}^{\infty } dp \frac{p^{2}}{e^{\frac{p-n}{T}}+1}&=\frac{1}{3}\pi^{3}nGT^{2}+2 ln2\cdot \pi n^{2}GT+3\pi \zeta\left ( 3 \right )GT^{3}, \nonumber\\
2 \pi G \int_{0}^{\infty } dp \frac{p^{2}}{e^{\frac{p+n}{T}}+1}&=-4 \pi GT^{3} Li_{3}\left ( -e^{-\frac{n}{T}} \right ). \nonumber
\end{align}

\section{The solution of $V_{\chi\mu}$}
In this appendix, we show how to get the solution eq.(\ref{eq25}). The transport equations for $V_{\chi\mu}$ to $\hbar$ order is
\begin{align}
\widetilde{p}_{\chi \mu }^{\left ( 0 \right )}V_{\chi }^{\left ( 1 \right )\mu }+J_{\chi \mu }^{\left ( 1 \right )}V_{\chi }^{\left ( 1 \right )\mu }=0, \nonumber\\
\bigtriangledown_{\chi \mu } V_{\chi }^{\left ( 1 \right )\mu }-\bigtriangledown J_{\chi \mu }^{\left ( 1 \right )}V_{\chi }^{\left ( 0 \right )\mu }=0, \nonumber\\
\varepsilon_{\mu \nu \sigma \rho } \bigtriangledown _{\chi }^{\sigma }V_{\chi }^{\left ( 0 \right )\rho }=-2\chi \left ( \widetilde{p}_{\chi \mu }^{\left ( 0 \right )}V_{\chi \nu }^{\left ( 1 \right )}-\widetilde{p}_{\chi \nu }^{\left ( 0 \right )}V_{\chi \mu }^{\left ( 1 \right )}+J_{\chi \mu }^{\left ( 1 \right )}V_{\chi \nu }^{\left ( 0 \right )}-J_{\chi \nu }^{\left ( 1 \right )}V_{\chi \mu }^{\left ( 0 \right )} \right ). \label{b1}
\end{align}
From the last equation of eq.(\ref{b1}), we can get
\begin{align}
\varepsilon_{\mu \nu \sigma \rho } \widetilde{p}_{\chi }^{\nu }\bigtriangledown ^{\sigma }_{\chi }V_{\chi }^{\left ( 0 \right )\rho }=4\chi \widetilde{p}_{\chi \mu }J_{\chi }^{\left ( 1 \right )\nu }V_{\chi \nu }^{\left ( 0 \right )}+2\chi \widetilde{p}_{\chi }^{2}V_{\chi \mu }^{\left ( 1 \right )}-2\chi J_{\chi \mu }^{\left ( 1 \right )}\widetilde{p}_{\chi }^{\nu }V_{\chi \nu }^{\left ( 0 \right )}, \label{b2}\\
V_{\chi \mu }^{\left ( 1 \right )}=\frac{\chi }{2\widetilde{p}^{2}_{\chi }}\varepsilon _{\mu \nu \sigma \rho }\widetilde{p}_{\chi }^{\nu }\bigtriangledown _{\chi }^{\sigma }V_{\chi }^{\left ( 0 \right )\rho }-\frac{2\left ( \widetilde{p}_{\chi }\cdot J_{\chi }^{\left ( 1 \right )} \right )}{\widetilde{p}^{2}_{\chi }}V_{\chi \mu }^{\left ( 0 \right )}+\frac{J^{\left ( 1 \right )}_{\chi \mu }\widetilde{p}^{\nu }_{\chi }V_{\chi \nu }^{\left ( 0 \right )}}{\widetilde{p}_{\chi }^{2}}. \label{b3}
\end{align}
So the solution of $V_{\chi\mu}$ to $\hbar$ order is 
\begin{align}
V_{\chi \mu }^{\left ( 1 \right )}=\widetilde{p}_{\chi \mu }f_{\chi }^{\left ( 1 \right )}\delta \left ( \widetilde{p}^{2} \right )+\kappa _{\mu }\delta \left ( \widetilde{p}_{\chi }^{2} \right )+\frac{\chi }{2\widetilde{p}^{2}_{\chi }}\varepsilon _{\mu \nu \sigma \rho }\widetilde{p}_{\chi }^{\nu }\bigtriangledown _{\chi }^{\sigma }V_{\chi }^{\left ( 0 \right )\rho }-\frac{2\left ( \widetilde{p}_{\chi }\cdot J_{\chi }^{\left ( 1 \right )} \right )}{\widetilde{p}^{2}_{\chi }}V_{\chi \mu }^{\left ( 0 \right )}+\frac{J^{\left ( 1 \right )}_{\chi \mu }\widetilde{p}^{\nu }_{\chi }V_{\chi \nu }^{\left ( 0 \right )}}{\widetilde{p}_{\chi }^{2}}. \label{b4}
\end{align}
Combining the last equation of eq.(\ref{b1}) and eq.(\ref{b4}), one can get
\begin{align}
\varepsilon_{\mu \nu \sigma \rho } \bigtriangledown ^{\sigma }_{\chi }V_{\chi }^{\left ( 0 \right )\rho }&=-2\chi\left [ \left ( \widetilde{p}_{\chi \mu }\kappa _{\nu }-\widetilde{p}_{\chi \nu }\kappa _{\mu } \right )\delta \left ( \widetilde{p}_{\chi }^{2} \right )+\frac{\chi }{2\widetilde{p}^{2}_{\chi }}\left ( \widetilde{p}_{\chi \mu }\varepsilon _{\nu \alpha \sigma \rho }-\widetilde{p}_{\chi \nu }\varepsilon _{\mu \alpha \sigma \rho } \right )\widetilde{p}_{\chi }^{\alpha }\bigtriangledown _{\chi }^{\sigma }V_{\chi }^{\left ( 0 \right )\rho } \right ] \nonumber\\
&-2\chi\left ( \widetilde{p}_{\chi \mu }J_{\chi \nu }^{\left ( 1 \right )}-\widetilde{p}_{\chi \nu }J_{\chi \mu }^{\left ( 1 \right )} \right )f_{\chi }^{\left ( 0 \right )}\delta \left ( \widetilde{p}^{2}_{\chi } \right )-2\chi \left ( J_{\chi \mu }^{\left ( 1 \right )}V_{\chi \nu }^{\left ( 0 \right )}-J_{\chi \nu }^{\left ( 1 \right )}V_{\chi \mu }^{\left ( 0 \right )} \right ), \label{b5}\\
\left ( \widetilde{p}_{\chi \mu }\kappa _{\nu }-\widetilde{p}_{\chi \nu }\kappa _{\mu } \right )\delta \left ( \widetilde{p}_{\chi }^{2} \right )&=-\frac{\chi }{2}\varepsilon _{\mu \nu \sigma \rho }\bigtriangledown _{\chi }^{\sigma }V_{\chi }^{\left ( 0 \right )\rho }-\frac{\chi }{2\widetilde{p}^{2}_{\chi }}\left ( \widetilde{p}_{\chi \mu }\varepsilon _{\nu \alpha \sigma \rho }-\widetilde{p}_{\chi \nu }\varepsilon _{\mu \alpha \sigma \rho } \right )\widetilde{p}^{\alpha }_{\chi }\bigtriangledown ^{\sigma }_{\chi }V_{\chi }^{\left ( 0 \right )\rho } \nonumber\\
&-\left ( J_{\chi \mu }^{\left ( 1 \right )}V_{\chi \nu }^{\left ( 0 \right )}-J_{\chi \nu }^{\left ( 1 \right )}V_{\chi \mu }^{\left ( 0 \right )} \right )-\left ( \widetilde{p}_{\chi \mu }J_{\chi \nu }^{\left ( 1 \right )}-\widetilde{p}_{\chi \nu }J_{\chi \mu }^{\left ( 1 \right )} \right )f_{\chi }^{\left ( 0 \right )}\delta \left ( \widetilde{p}^{2}_{\chi } \right ) \nonumber\\
&=-\frac{\chi }{2}\varepsilon _{\mu \nu \sigma \rho }\bigtriangledown _{\chi }^{\sigma }V_{\chi }^{\left ( 0 \right )\rho }-\frac{\chi }{2\widetilde{p}^{2}_{\chi }}\left ( \widetilde{p}_{\chi \mu }\varepsilon _{\nu \alpha \sigma \rho }-\widetilde{p}_{\chi \nu }\varepsilon _{\mu \alpha \sigma \rho } \right )\widetilde{p}^{\alpha }_{\chi }\bigtriangledown ^{\sigma }_{\chi }V_{\chi }^{\left ( 0 \right )\rho } \nonumber\\
&=-\frac{\chi }{2\widetilde{p}^{2}_{\chi }}\left ( \widetilde{p}_{\chi \alpha }\varepsilon _{\mu \nu \sigma \rho }+\widetilde{p}_{\chi \mu }\varepsilon _{\nu \alpha \sigma \rho }-\widetilde{p}_{\chi \nu }\varepsilon _{\mu \alpha \sigma \rho } \right )\widetilde{p}^{\alpha }_{\chi }\bigtriangledown ^{\sigma }_{\chi }V_{\chi }^{\left ( 0 \right )\rho } \nonumber\\
&=\frac{\chi }{2\widetilde{p}^{2}_{\chi }}\left ( \widetilde{p}_{\chi \sigma }\varepsilon _{\rho \alpha \mu \nu  }+\widetilde{p}_{\chi \rho }\varepsilon _{ \alpha \mu \nu \sigma  } \right )\widetilde{p}^{\alpha }_{\chi }\bigtriangledown ^{\sigma }_{\chi }V_{\chi }^{\left ( 0 \right )\rho } \nonumber\\
&=\frac{\chi }{2\widetilde{p}^{2}_{\chi }}\widetilde{p}_{\chi \rho}\widetilde{p}_{\chi }^{\alpha }\varepsilon _{\mu \nu \alpha \sigma }\left ( \bigtriangledown _{\chi }^{\sigma }V_{\chi }^{\left ( 0 \right )\rho }-\bigtriangledown _{\chi }^{\rho }V_{\chi }^{\left ( 0 \right )\sigma } \right ).\label{b6}
\end{align}
The derivatives of $V^{(0)}_\mu$ can be expanded
\begin{align}
\bigtriangledown _{\chi }^{\sigma }V_{\chi }^{\left ( 0 \right )\rho }&=F_{\chi }^{\sigma \rho }f_{\chi }^{\left ( 0 \right )}\delta \left ( \widetilde{p}^{2}_{\chi } \right )+\widetilde{p}^{\rho }_{\chi }\left ( \bigtriangledown ^{\sigma }_{\chi }f^{\left ( 0 \right )}_{\chi } \right )\delta \left ( \widetilde{p}^{2}_{\chi } \right )+2\widetilde{p}_{\chi }^{\rho }\widetilde{p}_{\chi \delta }F_{\chi }^{\sigma \delta }f_{\chi }^{\left ( 0 \right )}{}\ \delta '\left ( \widetilde{p}^{2}_{\chi } \right ),\label{b7} \\
\bigtriangledown _{\chi }^{\sigma }V_{\chi }^{\left ( 0 \right )\rho }-\bigtriangledown _{\chi }^{\rho }V_{\chi }^{\left ( 0 \right )\sigma }&=2F_{\chi }^{\sigma \rho }f_{\chi }^{\left ( 0 \right )}\delta \left ( \widetilde{p}^{2}_{\chi } \right )+\left [ \widetilde{p}_{\chi }^{\rho }\left ( \bigtriangledown ^{\sigma} _{\chi }f_{\chi }^{\left ( 0 \right )} \right )-\widetilde{p}_{\chi }^{\sigma }\left ( \bigtriangledown ^{\rho} _{\chi }f_{\chi }^{\left ( 0 \right )} \right ) \right ]\delta \left ( \widetilde{p}^{2}_{\chi } \right ) \nonumber\\
&+2\left ( \widetilde{p}^{\rho }_{\chi }\widetilde{p}_{\chi \delta }F^{\sigma \delta }_{\chi }-\widetilde{p}^{\sigma }_{\chi }\widetilde{p}_{\rho \delta }F^{\sigma \delta }_{\chi } \right ){}\ \delta '\left ( \widetilde{p}^{2}_{\chi } \right ). \label{b8}
\end{align}
So
\begin{align}
\left ( \widetilde{p}_{\chi \mu }\kappa _{\nu }-\widetilde{p}_{\chi \nu }\kappa _{\mu } \right )\delta \left ( \widetilde{p}_{\chi }^{2} \right )&=\frac{\chi }{2}\varepsilon _{\mu \nu \alpha  \sigma  }\widetilde{p}^{\alpha }_{\chi }\left ( \bigtriangledown _{\chi }^{\sigma }f_{\chi }^{\left ( 0 \right )} \right )\delta \left ( \widetilde{p}^{2}_{\chi } \right ). \label{b9}
\end{align}
Then we introduce the auxiliary unit vector $u'$ which is perpendicular to $\kappa_{\mu}$
\begin{align}
u^{'\mu}\kappa_{\mu}=0. \label{b10}
\end{align}
So we can get the solution of $\kappa_{\mu}\delta \left ( \widetilde{p}_{\chi }^{2} \right )$
\begin{align}
\kappa_{\mu}\delta \left ( \widetilde{p}_{\chi }^{2} \right )&=-\frac{\chi }{2\left ( \widetilde{p}_{\chi }\cdot u' \right )}\varepsilon _{\mu \nu \alpha \sigma }u^{'\nu }\widetilde{p}_{\chi }^{\alpha }\left ( \bigtriangledown ^{\sigma }_{\chi }f_{\chi }^{\left ( 0 \right )} \right )\delta \left ( \widetilde{p}_{\chi }^{2} \right ), \label{b11}
\end{align}
\begin{align}
V_{\chi \mu }^{\left ( 1 \right )}&=\widetilde{p}_{\chi \mu }f_{\chi }^{\left ( 1 \right )}\delta \left ( \widetilde{p}^{2}_{\chi } \right )-\frac{\chi }{2\left ( \widetilde{p}_{\chi }\cdot u' \right )}\varepsilon _{\mu \nu \alpha \sigma }u^{'\nu }\widetilde{p}_{\chi }^{\alpha }\left ( \bigtriangledown ^{\sigma }_{\chi }f_{\chi }^{\left ( 0 \right )} \right )\delta \left ( \widetilde{p}_{\chi }^{2} \right ) \nonumber\\
&+\frac{\chi }{2\widetilde{p}^{2}_{\chi }}\varepsilon _{\mu \nu \sigma \rho }\widetilde{p}_{\chi }^{\nu }\bigtriangledown _{\chi }^{\sigma }V_{\chi }^{\left ( 0 \right )\rho }-\frac{2\left ( \widetilde{p}_{\chi }\cdot J_{\chi }^{\left ( 1 \right )} \right )}{\widetilde{p}^{2}_{\chi }}V_{\chi \mu }^{\left ( 0 \right )}+\frac{J^{\left ( 1 \right )}_{\chi \mu }\widetilde{p}^{\nu }_{\chi }V_{\chi \nu }^{\left ( 0 \right )}}{\widetilde{p}_{\chi }^{2}}. \label{b12}
\end{align}

\section{The transport equation for $f_{\chi}$}
The transport equation is
\begin{align}
\bigtriangledown^{\mu } _{\chi }V_{\chi \mu }&=\left ( \partial^{\mu }-\partial^{\beta }J_{\chi }^{\left ( 0 \right )\mu }\partial ^{p}_{\beta } -\hbar \partial^{\beta }J_{\chi }^{\left ( 1 \right )\mu }\partial ^{p}_{\beta } \right )V_{\chi \mu }=0. \label{c1}
\end{align}
When we combine eq.(\ref{b13}) and eq.(\ref{c1}), to $\hbar$ order, the equation can be separated into six parts .\\
The first term is
\begin{align}
\left ( \partial^{\mu }-\partial^{\beta }J_{\chi }^{\left ( 0 \right )\mu }\partial ^{p}_{\beta } \right )\left [ \widetilde{p}_{\chi \mu }f_{\chi }\delta \left ( \widetilde{p}^{2}_{\chi } \right ) \right ]&= \widetilde{p}_{\chi \mu }\left( \bigtriangledown^{\mu}_{\chi} f_{\chi }\right)\delta \left ( \widetilde{p}^{2}_{\chi } \right ) . \label{c2}
\end{align}
The second term is
\begin{align}
&\frac{\chi \hbar}{2}\varepsilon _{\mu \nu \sigma \rho }\left ( \partial^{\mu }-\partial^{\beta }J_{\chi }^{\left ( 0 \right )\mu }\partial ^{p}_{\beta } \right )\left [ \frac{1}{\widetilde{p}^{2}_{\chi }}\widetilde{p}_{\chi }^{\nu }\Omega ^{\sigma \rho }_{\chi }f_{\chi }^{\left ( 0 \right )}\delta \left ( \widetilde{p}^{2}_{\chi } \right ) \right ] \nonumber\\
&=\chi\hbar\left ( \partial^{\mu }-\partial^{\beta }J_{\chi }^{\left ( 0 \right )\mu }\partial ^{p}_{\beta } \right )\left [ \frac{1}{\widetilde{p}^{2}_{\chi }}\widetilde{p}_{\chi }^{\nu }\widetilde{\Omega }_{\chi \mu \nu }f_{\chi }^{\left ( 0 \right )}\delta \left ( \widetilde{p}^{2}_{\chi } \right ) \right ] \nonumber\\
&=\chi \hbar\frac{\Omega ^{\mu \nu }_{\chi }\widetilde{p}^{2}-4\widetilde{p}^{\nu }\widetilde{p}_{\delta }\Omega ^{\mu \delta }_{\chi }}{\widetilde{p}^{4}}\widetilde{\Omega }_{\chi \mu \nu }f_{\chi }^{\left ( 0 \right )}\delta \left ( \widetilde{p}^{2}_{\chi } \right )+\frac{\chi\hbar}{\widetilde{p}^{2}_{\chi }}\widetilde{p}_{\chi }^{\nu }\left(\partial^{\mu }\widetilde{\Omega }_{\chi \mu \nu }\right)f_{\chi }^{\left ( 0 \right )}\delta \left ( \widetilde{p}^{2}_{\chi } \right ) \nonumber\\
&+\frac{\chi \hbar}{\widetilde{p}^{2}_{\chi }}\widetilde{p}^{\nu }_{\chi }\widetilde{\Omega }_{\chi \mu \nu }\left ( \bigtriangledown ^{\mu }_{\chi }f_{\chi }^{\left ( 0 \right )} \right )\delta \left ( \widetilde{p}^{2}_{\chi } \right )\nonumber\\
&=\frac{\chi \hbar}{\widetilde{p}^{2}_{\chi }}\widetilde{p}^{\nu }_{\chi }\widetilde{\Omega }_{\chi \mu \nu }\left ( \bigtriangledown ^{\mu }_{\chi }f_{\chi }^{\left ( 0 \right )} \right )\delta \left ( \widetilde{p}^{2}_{\chi } \right ). \label{c3}
\end{align}
The third term is
\begin{align}
&\left ( \partial^{\mu }-\partial^{\beta }J_{\chi }^{\left ( 0 \right )\mu }\partial ^{p}_{\beta } \right )\left [ \frac{\chi \hbar}{2\left ( \widetilde{p}_{\chi }\cdot u \right )}\varepsilon _{\mu \nu \sigma \rho }\widetilde{p}^{\alpha }_{\chi }u^{\nu }\left ( \bigtriangledown _{\chi }^{\sigma }f^{\left ( 0 \right )}_{\chi } \right )\delta \left ( \widetilde{p}^{2}_{\chi } \right ) \right ] \nonumber\\
&=\frac{\chi \hbar}{2\left ( \widetilde{p}_{\chi }\cdot u \right )^{2}}\varepsilon _{\mu \nu \alpha  \sigma }u_{\beta }u^{\nu }\widetilde{p}_{\chi }^{\alpha }\Omega _{\chi }^{\beta \mu }\left ( \bigtriangledown _{\chi }^{\sigma }f^{\left ( 0 \right )}_{\chi } \right )\delta \left ( \widetilde{p}^{2}_{\chi } \right ) \nonumber\\
&+\frac{\chi \hbar}{2\widetilde{p}_{\chi }\cdot u}\varepsilon _{\mu \nu \alpha \sigma }u^{\nu }\Omega ^{\mu \alpha }_{\chi }\left ( \bigtriangledown _{\chi }^{\sigma }f^{\left ( 0 \right )}_{\chi } \right )\delta \left ( \widetilde{p}^{2}_{\chi } \right ) \nonumber\\
&+\frac{\chi \hbar}{\widetilde{p}_{\chi }\cdot u}\varepsilon _{\mu \nu \alpha \sigma }u^{\nu }\widetilde{p}^{\alpha }_{\chi }\widetilde{p}_{\chi \beta }\Omega _{\chi }^{\mu \beta }\left ( \bigtriangledown _{\chi }^{\sigma }f^{\left ( 0 \right )}_{\chi } \right ){}\ \delta '\left ( \widetilde{p}^{2}_{\chi } \right ) \nonumber\\
&+\frac{\chi \hbar}{2\widetilde{p}_{\chi }\cdot u}\varepsilon _{\mu \nu \alpha \sigma }u^{\nu }\widetilde{p}_{\chi }^{\alpha }\left ( \bigtriangledown ^{\mu }_{\chi }\bigtriangledown ^{\sigma }_{\chi }f_{\chi }^{\left ( 0 \right )} \right )\delta \left ( \widetilde{p}^{2}_{\chi } \right ) \nonumber\\
&+\frac{\chi \hbar}{2\widetilde{p}_{\chi }\cdot u}\varepsilon _{\mu \nu \alpha \sigma }\left ( \partial ^{\mu }u^{\nu } \right )\widetilde{p}^{\alpha }_{\chi }\left ( \bigtriangledown _{\chi }^{\sigma }f^{\left ( 0 \right )}_{\chi } \right )\delta \left ( \widetilde{p}^{2}_{\chi } \right ) \nonumber\\
&-\frac{\chi \hbar}{2\left ( \widetilde{p}_{\chi }\cdot u \right )^{2}}\varepsilon _{\mu \nu \alpha  \sigma }u^{\nu}\widetilde{p}^{\alpha }_{\chi }\widetilde{p}_{\chi \beta }\left ( \partial ^{\mu }u^{\beta } \right )\left ( \bigtriangledown _{\chi }^{\sigma }f^{\left ( 0 \right )}_{\chi } \right )\delta \left ( \widetilde{p}^{2}_{\chi } \right ). \label{c4}
\end{align}
The fourth term is
\begin{align}
\left ( \partial^{\mu }-\partial^{\beta }J_{\chi }^{\left ( 0 \right )\mu }\partial ^{p}_{\beta } \right )\left [ \frac{2\hbar}{\widetilde{p}^{2}_{\chi }}\widetilde{p}^{\nu }_{\chi }\widetilde{p}_{\chi \mu }J^{\left ( 1 \right )}_{\chi \nu }f_{\chi }^{\left ( 0 \right )}\delta \left ( \widetilde{p}^{2}_{\chi } \right ) \right ]&=\frac{2\hbar}{\widetilde{p}^{4}_{\chi }}\left ( \Omega ^{\mu \nu }_{\chi }\widetilde{p}^{2}_{\chi }-4\Omega ^{\mu \delta }_{\chi }\widetilde{p}_{\chi \delta }\widetilde{p}_{\chi }^{\nu } \right )\widetilde{p}_{\chi \mu }J_{\chi \nu }^{\left ( 1 \right )}f_{\chi }^{\left ( 0 \right )}\delta \left ( \widetilde{p}^{2}_{\chi } \right ) \nonumber\\
&+\frac{2\hbar}{\widetilde{p}^{2}_{\chi }}\widetilde{p}^{\nu }_{\chi }\widetilde{p}_{\chi \mu }\bigtriangledown _{\chi }^{\mu }\left ( J_{\chi \nu }^{\left ( 1 \right )}f_{\chi }^{\left ( 0 \right )} \right )\delta \left ( \widetilde{p}^{2}_{\chi } \right ) \nonumber\\
&=\frac{2\hbar}{\widetilde{p}^{2}_{\chi }}\widetilde{p}^{\nu }_{\chi }\widetilde{p}_{\chi \mu }J_{\chi \nu }^{\left ( 1 \right )}\left ( \bigtriangledown _{\chi }^{\mu }f_{\chi }^{\left ( 0 \right )} \right )\delta \left ( \widetilde{p}^{2}_{\chi } \right ) \nonumber\\
&+\frac{2\hbar}{\widetilde{p}^{2}_{\chi }}\widetilde{p}^{\nu }_{\chi }\widetilde{p}_{\chi \mu }\left ( \partial ^{\mu }J_{\chi \nu }^{\left ( 1 \right )} \right )f_{\chi }^{\left ( 0 \right )}\delta \left ( \widetilde{p}^{2}_{\chi } \right ). \label{c5}
\end{align}
The fifth term is
\begin{align}
\left ( \partial^{\mu }-\partial^{\beta }J_{\chi }^{\left ( 0 \right )\mu }\partial ^{p}_{\beta } \right )\left [ \hbar J_{\chi \mu  }^{\left ( 1 \right )}f_{\chi }^{\left ( 0 \right )}\delta \left ( \widetilde{p}^{2}_{\chi } \right ) \right ]&=\hbar \left ( \partial ^{\mu }J_{\chi \mu }^{\left ( 1 \right )} \right )f_{\chi }^{\left ( 0 \right )}+\hbar J_{\chi \mu }^{\left ( 1 \right )}\left ( \bigtriangledown _{\chi }^{\mu }f_{\chi }^{\left ( 0 \right )} \right )\delta \left ( \widetilde{p}^{2}_{\chi } \right ) \nonumber\\
&+\hbar \widetilde{p}_{\chi \delta }\Omega _{\chi }^{\mu \delta }J_{\chi \mu }^{\left ( 1 \right )}f_{\chi }^{\left ( 0 \right )}{}\ \delta '\left ( \widetilde{p}^{2}_{\chi } \right ). \label{c6}
\end{align}
The sixth term is
\begin{align}
\hbar \partial ^{\beta }J_{\chi }^{\left ( 1 \right )\mu }\partial _{\beta }^{p}\left [ \widetilde{p}_{\chi \mu }f_{\chi }^{\left ( 0 \right )}\delta \left ( \widetilde{p}^{2}_{\chi } \right ) \right ]&=\hbar \left ( \partial _{\mu }J_{\chi }^{\left ( 1 \right )\mu } \right )f_{\chi }^{\left ( 0 \right )}\delta \left ( \widetilde{p}_{\chi }^{2} \right )+\hbar \partial ^{\beta }J_{\chi }^{\left ( 1 \right )\mu }\widetilde{p}_{\chi \mu }\left ( \partial _{\beta }^{p}f_{\chi }^{\left ( 0 \right )} \right )\delta \left ( \widetilde{p}_{\chi }^{2} \right ) \nonumber\\
&+2\hbar \partial ^{\beta }J_{\chi \mu }^{\left ( 1 \right )\mu }\widetilde{p}_{\chi \mu }\widetilde{p}_{\chi \beta }f_{\chi }^{\left ( 0 \right )}{}\ \delta '\left ( \widetilde{p}^{2}_{\chi } \right ). \label{c7}
\end{align}
We have the relation
\begin{align}
\varepsilon_{\mu \nu \alpha \sigma } \left ( \bigtriangledown _{\chi }^{\mu }\bigtriangledown _{\chi }^{\sigma }f_{\chi }^{\left ( 0 \right )} \right )=\frac{1}{2}\varepsilon _{\mu \nu \alpha \sigma }\left [ \bigtriangledown _{\chi }^{\mu } ,\bigtriangledown _{\chi }^{\sigma } \right ]f_{\chi }^{\left ( 0 \right )}=-\partial^{\beta } \widetilde{F}_{\chi }^{\nu \alpha }\partial _{\beta }^{p}f_{\chi }^{\left ( 0 \right )}, \label{c8}\\
2\widetilde{p}_{\chi }^{\beta }\varepsilon ^{\mu \nu \alpha \sigma }u_{\nu }\widetilde{p}_{\chi \alpha }\Omega _{\chi \mu \beta }=-2\left ( \widetilde{p}_{\chi }\cdot u \right )\widetilde{\Omega }_{\chi }^{\sigma \alpha }\widetilde{p}_{\chi \alpha }+2\widetilde{p}_{\chi }^{2}\widetilde{\Omega }_{\chi }^{\sigma \nu }u_{\nu }-2\widetilde{p}_{\chi \alpha }\widetilde{\Omega }_{\chi }^{\alpha \nu }u_{\nu }\widetilde{p}_{\chi }^{\sigma }.\label{c9}
\end{align}
So the transport equation can be written as
\begin{align}
&\left ( \partial^{\mu }-\partial^{\beta }J_{\chi }^{\left ( 0 \right )\mu }\partial ^{p}_{\beta } -\hbar \partial^{\beta }J_{\chi }^{\left ( 1 \right )\mu }\partial ^{p}_{\beta } \right )V_{\chi \mu } \nonumber\\
&=\widetilde{p}_{\chi \mu }\left( \bigtriangledown^{\mu}_{\chi} f_{\chi }\right)\delta \left ( \widetilde{p}^{2}_{\chi } \right )+\frac{\chi \hbar}{\widetilde{p}_{\chi }\cdot u}\widetilde{p}_{\chi }^{\alpha }\widetilde{\Omega }_{\chi \alpha \nu }u^{\nu }\widetilde{p}_{\chi \sigma }\left ( \bigtriangledown _{\chi }^{\sigma }f_{\chi }^{\left ( 0 \right )} \right ){}\ \delta '\left ( \widetilde{p}^{2}_{\chi } \right ) \nonumber\\
&+2\hbar\widetilde{p}_{\chi \nu }J_{\chi }^{\left ( 0 \right )\nu }\widetilde{p}_{\chi \mu }\left ( \bigtriangledown _{\chi }^{\mu }f_{\chi }^{\left ( 0 \right )} \right ){}\ \delta '\left ( \widetilde{p}^{2}_{\chi } \right ) \nonumber\\
&+\frac{\chi \hbar}{2\left ( \widetilde{p}_{\chi }\cdot u \right )^{2}}\left [ \left ( \partial ^{\mu }u^{\beta }\right)\widetilde{p}_{\chi \beta }-\Omega _{\chi }^{\beta \mu }u_{\beta }  \right ]\varepsilon _{\mu \nu \alpha \sigma }u^{\nu }\widetilde{p}_{\chi }^{\alpha }\left ( \bigtriangledown _{\chi }^{\sigma }f_{\chi }^{\left ( 0 \right )} \right )\delta \left ( \widetilde{p}_{\chi }^{2} \right ) \nonumber\\
&-\frac{\chi \hbar}{2\widetilde{p}_{\chi }\cdot u}\varepsilon_{\mu \nu \alpha \sigma }\left ( \partial ^{\mu }u^{\nu } \right ) \widetilde{p}_{\chi }^{\alpha }\left ( \bigtriangledown _{\chi }^{\sigma }f_{\chi }^{\left ( 0 \right )} \right )\delta \left ( \widetilde{p}_{\chi }^{2} \right )+\frac{\chi \hbar}{2\widetilde{p}_{\chi }\cdot u}u^{\nu}\widetilde{p}_{\chi}^{\alpha}\left ( \partial ^{\beta }\widetilde{\Omega }_{\chi }^{\nu \alpha } \right )\left ( \partial _{\beta }^{p}f_{\chi }^{\left ( 0 \right )} \right )\delta \left ( \widetilde{p}_{\chi }^{2} \right ) \nonumber\\
&+\hbar J_{\chi}^{\left ( 1 \right )}\left ( \bigtriangledown _{\chi }^{\mu }f_{\chi }^{\left ( 0 \right )} \right )\delta \left ( \widetilde{p}_{\chi }^{2} \right )-\frac{\hbar}{\widetilde{p}_{\chi }^{2}}\widetilde{p}_{\chi \delta }\Omega _{\chi }^{\mu \delta }J_{\chi \mu }^{\left ( 1 \right )}f_{\chi }^{\left ( 0 \right )}\delta \left ( \widetilde{p}_{\chi }^{2} \right )-\hbar\partial^{\beta } J_{\chi }^{\left ( 1 \right )\mu }\widetilde{p}_{\chi \mu }\left ( \partial ^{p}_{\beta }f_{\chi } ^{\left ( 0 \right )}\right )\delta \left ( \widetilde{p}_{\chi }^{2} \right ) \nonumber\\
&=\delta\left ( \widetilde{p}_{\chi }^{2}+\frac{\chi \hbar}{\widetilde{p}_{\chi }\cdot u}\widetilde{p}_{\chi }^{\alpha }\widetilde{\Omega }_{\chi \alpha \nu }u^{\nu }+2\hbar\widetilde{p}_{\chi \nu }J_{\chi }^{\left ( 1 \right )\nu } \right ) \{ \widetilde{p}_{\chi }\cdot \bigtriangledown _{\chi }+\frac{\chi \hbar}{2\left ( \widetilde{p}_{\chi }\cdot u \right )^{2}}\left [ \left ( \partial ^{\mu }u^{\beta }\right)\widetilde{p}_{\chi \beta }-\Omega _{\chi }^{\beta \mu }u_{\beta }  \right ]\varepsilon _{\mu \nu \alpha \sigma }u^{\nu }\widetilde{p}_{\chi }^{\alpha } \bigtriangledown _{\chi }^{\sigma }\nonumber\\
&-\frac{\chi \hbar}{2\widetilde{p}_{\chi }\cdot u}\varepsilon_{\mu \nu \alpha \sigma }\left ( \partial ^{\mu }u^{\nu } \right ) \widetilde{p}_{\chi }^{\alpha }\bigtriangledown _{\chi }^{\sigma }+\frac{\chi \hbar}{2\widetilde{p}_{\chi }\cdot u}u^{\nu}\widetilde{p}_{\chi}^{\alpha}\left ( \partial ^{\beta }\widetilde{\Omega }_{\chi }^{\nu \alpha } \right )\partial _{\beta }^{p}+\hbar J_{\chi}^{\left ( 1 \right )}\bigtriangledown _{\chi }^{\mu }-\frac{\hbar}{\widetilde{p}_{\chi }^{2}}\widetilde{p}_{\chi \delta }\Omega _{\chi }^{\mu \delta }J_{\chi \mu }^{\left ( 1 \right )} \nonumber\\
&-\hbar\partial^{\beta } J_{\chi }^{\left ( 1 \right )\mu }\widetilde{p}_{\chi \mu }\partial ^{p}_{\beta } \}f_{\chi}=0. \label{c10}
\end{align}


\begin{thebibliography}{99}

%\cite{STAR:2021beb}
\bibitem{STAR:2021beb}
M.~S.~Abdallah \textit{et al.} [STAR],
%``Global $\Lambda$-hyperon polarization in Au+Au collisions at $\sqrt {s_{NN}}$=3~GeV,''
Phys. Rev. C \textbf{104} (2021) no.6, L061901
doi:10.1103/PhysRevC.104.L061901
[arXiv:2108.00044 [nucl-ex]].
%38 citations counted in INSPIRE as of 10 Oct 2022

%\cite{STAR:2019fge}
\bibitem{STAR:2019fge}
J.~Adam \textit{et al.} [STAR],
%``Measurement of inclusive $J/\psi$ suppression in Au+Au collisions at $\sqrt{s_{NN}}$ = 200 GeV through the dimuon channel at STAR,''
Phys. Lett. B \textbf{797} (2019), 134917
doi:10.1016/j.physletb.2019.134917
[arXiv:1905.13669 [nucl-ex]].
%29 citations counted in INSPIRE as of 10 Oct 2022

%\cite{CMS:2018zza}
\bibitem{CMS:2018zza}
A.~M.~Sirunyan \textit{et al.} [CMS],
%``Measurement of nuclear modification factors of $\Upsilon$(1S), $\Upsilon$(2S), and $\Upsilon$(3S) mesons in PbPb collisions at $\sqrt{s_{_\mathrm{NN}}} =$ 5.02 TeV,''
Phys. Lett. B \textbf{790} (2019), 270-293
doi:10.1016/j.physletb.2019.01.006
[arXiv:1805.09215 [hep-ex]].
%135 citations counted in INSPIRE as of 10 Oct 2022

%\cite{Kharzeev:2007jp}
\bibitem{Kharzeev:2007jp}
D.~E.~Kharzeev, L.~D.~McLerran and H.~J.~Warringa,
%``The Effects of topological charge change in heavy ion collisions: 'Event by event P and CP violation',''
Nucl. Phys. A \textbf{803} (2008), 227-253
doi:10.1016/j.nuclphysa.2008.02.298
[arXiv:0711.0950 [hep-ph]].
%1613 citations counted in INSPIRE as of 05 Sep 2022


%\cite{Son:2009tf}
\bibitem{Son:2009tf}
D.~T.~Son and P.~Surowka,
%``Hydrodynamics with Triangle Anomalies,''
Phys. Rev. Lett. \textbf{103} (2009), 191601
doi:10.1103/PhysRevLett.103.191601
[arXiv:0906.5044 [hep-th]].
%810 citations counted in INSPIRE as of 05 Sep 2022

%\cite{STAR:2023nqc}
\bibitem{STAR:2023nqc}
 [STAR],
%``Event-by-event correlations between $\Lambda$ ($\bar{\Lambda}$) hyperon global polarization and handedness with charged hadron azimuthal separation in Au+Au collisions at $\sqrt{s_{\text{NN}}} = 27 \text{ GeV}$ from STAR,''
[arXiv:2304.10037 [nucl-ex]].
%0 citations counted in INSPIRE as of 05 Jun 2023

%\cite{Sarkar:2022uyt}
\bibitem{Sarkar:2022uyt}
D.~Sarkar [ALICE],
%``Global and local polarization of $\Lambda (\overline{\Lambda})$ hyperons in Pb\textendash{}Pb collisions in ALICE at the LHC,''
EPJ Web Conf. \textbf{259} (2022), 06001
doi:10.1051/epjconf/202225906001
[arXiv:2209.04798 [hep-ex]].
%0 citations counted in INSPIRE as of 05 Jun 2023


%\cite{Gou:2023dkp}
\bibitem{Gou:2023dkp}
X.~Gou [STAR],
%``Measurements of global and local polarization of hyperons in isobar collisions at 200 GeV from STAR,''
EPJ Web Conf. \textbf{276} (2023), 04007
doi:10.1051/epjconf/202327604007
%0 citations counted in INSPIRE as of 05 Jun 2023


%\cite{Son:2012zy}
\bibitem{Son:2012zy}
D.~T.~Son and N.~Yamamoto,
%``Kinetic theory with Berry curvature from quantum field theories,''
Phys. Rev. D \textbf{87} (2013) no.8, 085016
doi:10.1103/PhysRevD.87.085016
[arXiv:1210.8158 [hep-th]].
%293 citations counted in INSPIRE as of 05 Sep 2022

%\cite{Gao:2012ix}
\bibitem{Gao:2012ix}
J.~H.~Gao, Z.~T.~Liang, S.~Pu, Q.~Wang and X.~N.~Wang,
%``Chiral Anomaly and Local Polarization Effect from Quantum Kinetic Approach,''
Phys. Rev. Lett. \textbf{109} (2012), 232301
doi:10.1103/PhysRevLett.109.232301
[arXiv:1203.0725 [hep-ph]].
%217 citations counted in INSPIRE as of 05 Sep 2022


%\cite{Chen:2012ca}
\bibitem{Chen:2012ca}
J.~W.~Chen, S.~Pu, Q.~Wang and X.~N.~Wang,
%``Berry Curvature and Four-Dimensional Monopoles in the Relativistic Chiral Kinetic Equation,''
Phys. Rev. Lett. \textbf{110} (2013) no.26, 262301
doi:10.1103/PhysRevLett.110.262301
[arXiv:1210.8312 [hep-th]].
%247 citations counted in INSPIRE as of 05 Sep 2022


%\cite{Huang:2018wdl}
\bibitem{Huang:2018wdl}
A.~Huang, S.~Shi, Y.~Jiang, J.~Liao and P.~Zhuang,
%``Complete and Consistent Chiral Transport from Wigner Function Formalism,''
Phys. Rev. D \textbf{98} (2018) no.3, 036010
doi:10.1103/PhysRevD.98.036010
[arXiv:1801.03640 [hep-th]].
%109 citations counted in INSPIRE as of 05 Sep 2022

%\cite{Wang:2020dws}
\bibitem{Wang:2020dws}
Z.~Wang, X.~Guo, S.~Shi and P.~Zhuang,
%``Mass Correction to Chiral Kinetic Equations,''
Nucl. Phys. A \textbf{1005} (2021), 121976
doi:10.1016/j.nuclphysa.2020.121976
[arXiv:2004.12174 [hep-ph]].
%3 citations counted in INSPIRE as of 05 Sep 2022

%\cite{Chen:2013iga}
\bibitem{Chen:2013iga}
J.~W.~Chen, J.~y.~Pang, S.~Pu and Q.~Wang,
%``Kinetic equations for massive Dirac fermions in electromagnetic field with non-Abelian Berry phase,''
Phys. Rev. D \textbf{89} (2014) no.9, 094003
doi:10.1103/PhysRevD.89.094003
[arXiv:1312.2032 [hep-th]].
%60 citations counted in INSPIRE as of 05 Sep 2022

%\cite{Gao:2019znl}
\bibitem{Gao:2019znl}
J.~H.~Gao and Z.~T.~Liang,
%``Relativistic Quantum Kinetic Theory for Massive Fermions and Spin Effects,''
Phys. Rev. D \textbf{100} (2019) no.5, 056021
doi:10.1103/PhysRevD.100.056021
[arXiv:1902.06510 [hep-ph]].
%96 citations counted in INSPIRE as of 05 Sep 2022

%\cite{Weickgenannt:2019dks}
\bibitem{Weickgenannt:2019dks}
N.~Weickgenannt, X.~L.~Sheng, E.~Speranza, Q.~Wang and D.~H.~Rischke,
%``Kinetic theory for massive spin-1/2 particles from the Wigner-function formalism,''
Phys. Rev. D \textbf{100} (2019) no.5, 056018
doi:10.1103/PhysRevD.100.056018
[arXiv:1902.06513 [hep-ph]].
%128 citations counted in INSPIRE as of 05 Sep 2022

%\cite{Hattori:2019ahi}
\bibitem{Hattori:2019ahi}
K.~Hattori, Y.~Hidaka and D.~L.~Yang,
%``Axial Kinetic Theory and Spin Transport for Fermions with Arbitrary Mass,''
Phys. Rev. D \textbf{100} (2019) no.9, 096011
doi:10.1103/PhysRevD.100.096011
[arXiv:1903.01653 [hep-ph]].
%98 citations counted in INSPIRE as of 05 Sep 2022

%\cite{Zamanian:2009jzf}
\bibitem{Zamanian:2009jzf}
J.~Zamanian, M.~Marklund and G.~Brodin,
%``Scalar quantum kinetic theory for spin-1/2 particles: Mean field theory,''
New J. Phys. \textbf{12} (2010), 043019
doi:10.1088/1367-2630/12/4/043019
[arXiv:0910.5165 [cond-mat.quant-gas]].
%17 citations counted in INSPIRE as of 05 Sep 2022

%\cite{Lin:2022tma}
\bibitem{Lin:2022tma}
S.~Lin and Z.~Wang,
%``Shear induced polarization: Collisional contributions,''
[arXiv:2206.12573 [hep-ph]].
%0 citations counted in INSPIRE as of 13 Oct 2022

%\cite{Liu:2022zxd}
\bibitem{Liu:2022zxd}
S.~Y.~F.~Liu, B.~Fu and L.~Pang,
%``Shear-induced spin polarization and \textquotedblleft{}strange memory\textquotedblright{} in heavy-ion collisions,''
EPJ Web Conf. \textbf{259} (2022), 13004
doi:10.1051/epjconf/202225913004
%0 citations counted in INSPIRE as of 13 Oct 2022

%\cite{Buzzegoli:2022fxu}
\bibitem{Buzzegoli:2022fxu}
M.~Buzzegoli, F.~Becattini, G.~Inghirami, I.~Karpenko and A.~Palermo,
%``Spin-thermal shear coupling in relativistic nuclear collisions,''
[arXiv:2208.04449 [nucl-th]].
%0 citations counted in INSPIRE as of 13 Oct 2022

%\cite{Becattini:2021suc}
\bibitem{Becattini:2021suc}
F.~Becattini, M.~Buzzegoli and A.~Palermo,
%``Spin-thermal shear coupling in a relativistic fluid,''
Phys. Lett. B \textbf{820} (2021), 136519
doi:10.1016/j.physletb.2021.136519
[arXiv:2103.10917 [nucl-th]].
%58 citations counted in INSPIRE as of 13 Oct 2022

%\cite{Wang:2021owk}
\bibitem{Wang:2021owk}
Z.~Wang and P.~Zhuang,
%``Spin Polarization Induced by Inhomogeneous Dynamical Condensate,''
[arXiv:2101.00586 [hep-ph]].
%6 citations counted in INSPIRE as of 26 Nov 2022

%\cite{Wang:2020pej}
\bibitem{Wang:2020pej}
Z.~Wang, X.~Guo and P.~Zhuang,
%``Equilibrium Spin Distribution From Detailed Balance,''
Eur. Phys. J. C \textbf{81} (2021) no.9, 799
doi:10.1140/epjc/s10052-021-09586-8
[arXiv:2009.10930 [hep-th]].
%38 citations counted in INSPIRE as of 26 Nov 2022


\bibitem{PhysRev.109.193}
R. P. Feynman and M. Gell-Mann,
Phys. Rev. \textbf{109} (1958) 1, 193-198
doi:10.1103/PhysRev.109.193



%\cite{Fukushima:2010zza}
\bibitem{Fukushima:2010zza}
K.~Fukushima and M.~Ruggieri,
%``Dielectric correction to the Chiral Magnetic Effect,''
Phys. Rev. D \textbf{82} (2010), 054001
doi:10.1103/PhysRevD.82.054001
[arXiv:1004.2769 [hep-ph]].
%39 citations counted in INSPIRE as of 06 Dec 2022

%\cite{Diakonov:2011fs}
\bibitem{Diakonov:2011fs}
D.~Diakonov, A.~G.~Tumanov and A.~A.~Vladimirov,
%``Low-energy General Relativity with torsion: A Systematic derivative expansion,''
Phys. Rev. D \textbf{84} (2011), 124042
doi:10.1103/PhysRevD.84.124042
[arXiv:1104.2432 [hep-th]].
%45 citations counted in INSPIRE as of 06 Dec 2022

%\cite{Andrianov:2002iw}
\bibitem{Andrianov:2002iw}
V.~A.~Andrianov and S.~S.~Afonin,
%``Vector particles in quasilocal quark models,''
Zap. Nauchn. Semin. \textbf{291} (2002), 5
[arXiv:hep-ph/0304140 [hep-ph]].
%11 citations counted in INSPIRE as of 06 Dec 2022

%\cite{Braun:2011pp}
\bibitem{Braun:2011pp}
J.~Braun,
%``Fermion Interactions and Universal Behavior in Strongly Interacting Theories,''
J. Phys. G \textbf{39} (2012), 033001
doi:10.1088/0954-3899/39/3/033001
[arXiv:1108.4449 [hep-ph]].
%284 citations counted in INSPIRE as of 06 Dec 2022





%\cite{Satow:2014lia}
\bibitem{Satow:2014lia}
D.~Satow,
%``Nonlinear electromagnetic response in quark-gluon plasma,''
Phys. Rev. D \textbf{90} (2014) no.3, 034018
doi:10.1103/PhysRevD.90.034018
[arXiv:1406.7032 [hep-ph]].
%26 citations counted in INSPIRE as of 27 Nov 2022

%\cite{Gorbar:2017toh}
\bibitem{Gorbar:2017toh}
E.~V.~Gorbar, D.~O.~Rybalka and I.~A.~Shovkovy,
%``Second-order dissipative hydrodynamics for plasma with chiral asymmetry and vorticity,''
Phys. Rev. D \textbf{95} (2017) no.9, 096010
doi:10.1103/PhysRevD.95.096010
[arXiv:1702.07791 [hep-th]].
%19 citations counted in INSPIRE as of 27 Nov 2022

%\cite{Buzzegoli:2017cqy}
\bibitem{Buzzegoli:2017cqy}
M.~Buzzegoli, E.~Grossi and F.~Becattini,
%``General equilibrium second-order hydrodynamic coefficients for free quantum fields,''
JHEP \textbf{10} (2017), 091
[erratum: JHEP \textbf{07} (2018), 119]
doi:10.1007/JHEP10(2017)091
[arXiv:1704.02808 [hep-th]].
%58 citations counted in INSPIRE as of 27 Nov 2022

%\cite{Buzzegoli:2018wpy}
\bibitem{Buzzegoli:2018wpy}
M.~Buzzegoli and F.~Becattini,
%``General thermodynamic equilibrium with axial chemical potential for the free Dirac field,''
JHEP \textbf{12} (2018), 002
[erratum: JHEP \textbf{03} (2022), 045]
doi:10.1007/JHEP12(2018)002
[arXiv:1807.02071 [hep-th]].
%32 citations counted in INSPIRE as of 27 Nov 2022

%\cite{Lin:2021sjw}
\bibitem{Lin:2021sjw}
S.~Lin and L.~Yang,
%``Magneto-vortical effect in strong magnetic field,''
JHEP \textbf{06} (2021), 054
doi:10.1007/JHEP06(2021)054
[arXiv:2103.11577 [nucl-th]].
%6 citations counted in INSPIRE as of 27 Nov 2022

%\cite{Florkowski:2018fap}
\bibitem{Florkowski:2018fap}
W.~Florkowski, A.~Kumar and R.~Ryblewski,
%``Relativistic hydrodynamics for spin-polarized fluids,''
Prog. Part. Nucl. Phys. \textbf{108} (2019), 103709
doi:10.1016/j.ppnp.2019.07.001
[arXiv:1811.04409 [nucl-th]].
%120 citations counted in INSPIRE as of 27 Nov 2022

%\cite{Bhadury:2020puc}
\bibitem{Bhadury:2020puc}
S.~Bhadury, W.~Florkowski, A.~Jaiswal, A.~Kumar and R.~Ryblewski,
%``Relativistic dissipative spin dynamics in the relaxation time approximation,''
Phys. Lett. B \textbf{814} (2021), 136096
doi:10.1016/j.physletb.2021.136096
[arXiv:2002.03937 [hep-ph]].
%82 citations counted in INSPIRE as of 27 Nov 2022

%\cite{Peng:2021ago}
\bibitem{Peng:2021ago}
H.~H.~Peng, J.~J.~Zhang, X.~L.~Sheng and Q.~Wang,
%``Ideal Spin Hydrodynamics from the Wigner Function Approach,''
Chin. Phys. Lett. \textbf{38} (2021) no.11, 116701
doi:10.1088/0256-307X/38/11/116701
[arXiv:2107.00448 [hep-th]].
%29 citations counted in INSPIRE as of 27 Nov 2022




%\cite{Guo:2017dzf}
\bibitem{Guo:2017dzf}
X.~Guo and P.~Zhuang,
%``Out-of-equilibrium $U_A$(1) symmetry breaking in electromagnetic fields,''
Phys. Rev. D \textbf{98} (2018) no.1, 016007
doi:10.1103/PhysRevD.98.016007
[arXiv:1711.02924 [hep-th]].
%13 citations counted in INSPIRE as of 06 Sep 2022

%\cite{Huang:2018wdl}
\bibitem{Huang:2018wdl}
A.~Huang, S.~Shi, Y.~Jiang, J.~Liao and P.~Zhuang,
%``Complete and Consistent Chiral Transport from Wigner Function Formalism,''
Phys. Rev. D \textbf{98} (2018) no.3, 036010
doi:10.1103/PhysRevD.98.036010
[arXiv:1801.03640 [hep-th]].
%115 citations counted in INSPIRE as of 02 Dec 2022


%\cite{Guo:2020zpa}
\bibitem{Guo:2020zpa}
X.~Guo,
%``Massless Limit of Transport Theory for Massive Fermions,''
Chin. Phys. C \textbf{44} (2020) no.10, 104106
doi:10.1088/1674-1137/ababf9
[arXiv:2005.00228 [hep-ph]].
%16 citations counted in INSPIRE as of 06 Sep 2022


%\cite{Lin:2021mvw}
\bibitem{Lin:2021mvw}
S.~Lin,
%``Quantum kinetic theory for quantum electrodynamics,''
Phys. Rev. D \textbf{105} (2022) no.7, 076017
doi:10.1103/PhysRevD.105.076017
[arXiv:2109.00184 [hep-ph]].
%17 citations counted in INSPIRE as of 30 Nov 2022

%\cite{Huang:2020wrr}
\bibitem{Huang:2020wrr}
A.~Huang, S.~Shi, X.~Zhu, L.~He, J.~Liao and P.~Zhuang,
%``Quantum kinetic equation and dynamical mass generation in 2+1 dimensions,''
Phys. Rev. D \textbf{103} (2021) no.5, 056025
doi:10.1103/PhysRevD.103.056025
[arXiv:2007.02858 [hep-th]].
%11 citations counted in INSPIRE as of 02 Dec 2022

%\cite{Wang:2017vtn}
\bibitem{Wang:2017vtn}
Z.~Wang and P.~Zhuang,
%``Meson properties in magnetized quark matter,''
Phys. Rev. D \textbf{97} (2018) no.3, 034026
doi:10.1103/PhysRevD.97.034026
[arXiv:1712.00554 [hep-ph]].
%39 citations counted in INSPIRE as of 02 Dec 2022



%\cite{Gorbar:2017awz}
\bibitem{Gorbar:2017awz}
E.~V.~Gorbar, V.~A.~Miransky, I.~A.~Shovkovy and P.~O.~Sukhachov,
%``Wigner function and kinetic phenomena for chiral plasma in a strong magnetic field,''
JHEP \textbf{08} (2017), 103
doi:10.1007/JHEP08(2017)103
[arXiv:1707.01105 [hep-ph]].
%16 citations counted in INSPIRE as of 13 Oct 2022

%\cite{Wu:2022mkr}
\bibitem{Wu:2022mkr}
X.~Y.~Wu, C.~Yi, G.~Y.~Qin and S.~Pu,
%``Local and global polarization of \ensuremath{\Lambda} hyperons across RHIC-BES energies: The roles of spin hall effect, initial condition, and baryon diffusion,''
Phys. Rev. C \textbf{105} (2022) no.6, 064909
doi:10.1103/PhysRevC.105.064909
[arXiv:2204.02218 [hep-ph]].
%15 citations counted in INSPIRE as of 30 Nov 2022



%\cite{Niida:2018hfw}
\bibitem{Niida:2018hfw}
T.~Niida [STAR],
%``Global and local polarization of $\Lambda$ hyperons in Au+Au collisions at 200 GeV from STAR,''
Nucl. Phys. A \textbf{982} (2019), 511-514
doi:10.1016/j.nuclphysa.2018.08.034
[arXiv:1808.10482 [nucl-ex]].
%69 citations counted in INSPIRE as of 30 Nov 2022



%\cite{Hidaka:2022dmn}
\bibitem{Hidaka:2022dmn}
Y.~Hidaka, S.~Pu, Q.~Wang and D.~L.~Yang,
%``Foundations and applications of quantum kinetic theory,''
Prog. Part. Nucl. Phys. \textbf{127} (2022), 103989
doi:10.1016/j.ppnp.2022.103989
[arXiv:2201.07644 [hep-ph]].
%19 citations counted in INSPIRE as of 26 Nov 2022


\end{thebibliography}
\end{document}